\documentclass[aps,prl,reprint,superscriptaddress]{revtex4-2}

\usepackage{graphicx}
\usepackage{siunitx}              % unit spacings, error typesetting, ...
\usepackage[T1]{fontenc}          % font encoding for special characters + embedding font in pdf (?)
\usepackage{lmodern}              % use vector fonts
\usepackage{amsmath}              % align n' stuff
\usepackage{bm}                   % bold math
\usepackage{hyperref}             % links

\hypersetup{hidelinks, colorlinks=false}

\newcommand{\mf}{m_\text{F}}
\newcommand{\phil}{\varphi_\text{L}}

\makeatletter
\let\cat@comma@active\@empty
\makeatother

\begin{document}
  \title{Collisions of three-component vector solitons in Bose-Einstein condensates}
  
  \author{Stefan Lannig}
  \email{vectorsolitons@matterwave.de}
  \author{Christian-Marcel Schmied}
  \author{Maximilian Pr\"ufer}
  \author{Philipp Kunkel}
  \author{Robin Strohmaier}
  \author{Helmut Strobel}
  \author{Thomas Gasenzer}
  \affiliation{Kirchhoff-Institut f\"ur Physik, Universit\"at Heidelberg, Im Neuenheimer Feld 227, 69120 Heidelberg, Germany}
  
  \author{Panayotis G. Kevrekidis}
  \affiliation{Department of Mathematics and Statistics, University of Massachusetts, Amherst, Massachusetts 01003-4515 USA}
  
  \author{Markus K. Oberthaler}
  \affiliation{Kirchhoff-Institut f\"ur Physik, Universit\"at Heidelberg, Im Neuenheimer Feld 227, 69120 Heidelberg, Germany}
  
  \date{\today}
  
  \begin{abstract}
    Ultracold gases provide an unprecedented level of control for the investigation of soliton dynamics and collisions.
    We present a scheme for deterministically preparing pairs of three-component solitons in a Bose-Einstein condensate. Our method is based on local spin rotations which simultaneously imprint suitable phase and density distributions.
    This enables us to observe striking collisional properties of the vector degree of freedom which naturally arises for the coherent nature of the emerging multi-component solitons.
    We find that the solitonic properties in the quasi-one-dimensional system are quantitatively described by the integrable repulsive three-component Manakov model.
  \end{abstract}
  
  \maketitle
  
  Solitons, non-dispersive wave packets in nonlinear systems, are realized in a broad variety of settings across nature -- from optics and classical fluids to plasmas and ultracold atoms \cite{kivshar2003,ablowitz2011}. While an extensive effort has been invested in the understanding of single-component systems, the study of coupled multi-component nonlinear models is far less developed, especially in settings involving more than two components.
  In the presence of well-defined phase relations between the constituent fields the concept of vector solitons arises and their internal degree of freedom leads to striking interaction features \cite{manakov1974}.
  
  There are different platforms for investigating solitonic collisions, most notably nonlinear optics systems where polarization shifts have been demonstrated \cite{mollenauer1995,anastassiou1999,anastassiou2001}.
  Nowadays, the unprecedented level of control available in ultracold atomic systems offers new perspectives.
  These systems do not only provide a variable number of internal states with long coherence times but also a wide variety of methods for manipulating and detecting the constituent fields.
  Single-component collisions have already been studied in great detail \cite{weller2008,stellmer2008,nguyen2014}.
  Recently, this has been extended to the experimental detection of two-component \cite{becker2008} and magnetic solitons \cite{chai2019,farolfi2019}, as well as three-component solitonic structures \cite{bersano2018}.
  
  \begin{figure}
    \includegraphics[width=\linewidth]{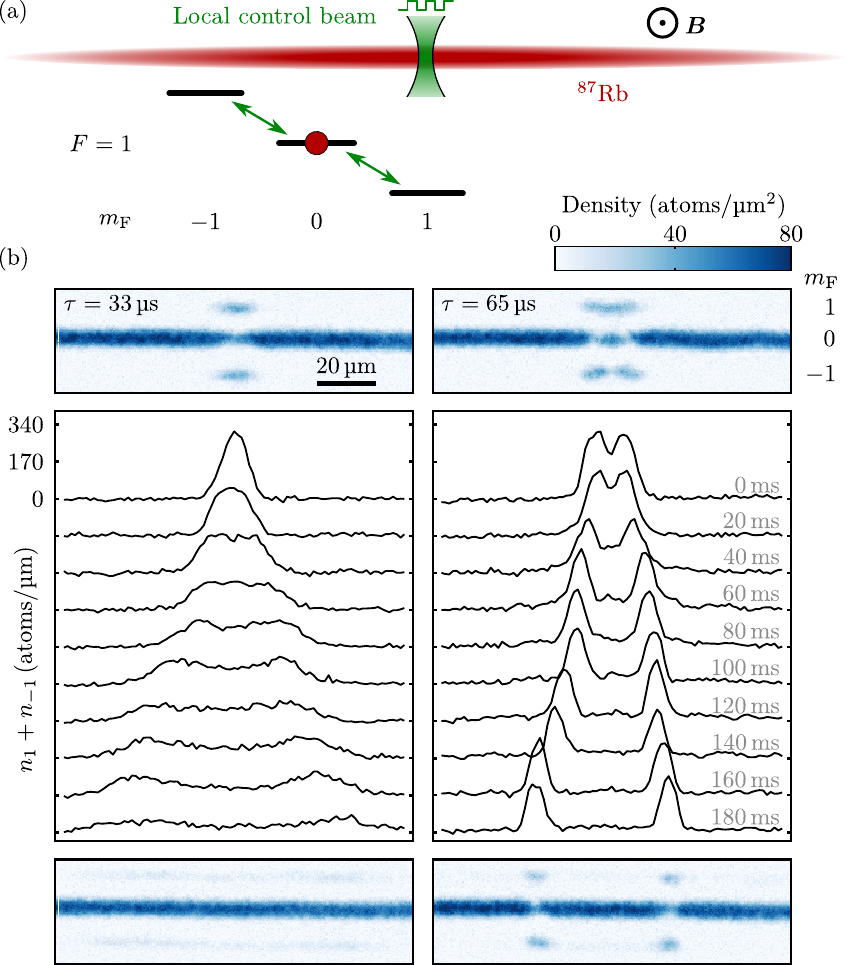}
    \caption{\label{fig:soliton_prep} 
      Formation of three-component vector solitons.
      (a) An amplitude-modulated steerable laser beam (green) is used to implement local spin rotations in an elongated BEC (red) subject to a homogeneous magnetic field $\bm{B}$ along the $z$-direction. This coherently transfers atoms from the initial state $\mf=0$ (red disk) to $\mf=\pm1$ (green arrows).
      (b) The upper panel shows absorption images after Stern-Gerlach separation revealing the density distributions of the three $\mf$ states after local spin rotations of duration $\tau$. For short Rabi coupling ($\tau=\SI{33}{\micro\second}$) the population is transferred to $\mf=\pm1$; and for longer pulse duration ($\tau=\SI{65}{\micro\second}$) the population is coherently transferred back to $\mf=0$ in the center of the laser beam, implying a sign change in $\psi_0$. The subsequent dynamics shown below (summed $\mf=\pm1$ densities $n_{\pm1}$) indicates the formation of a soliton pair as a result of the sign change (right column). Each soliton consists of shape-preserving bright components ($\mf=\pm1$) and a corresponding density depletion in the $\mf=0$ component (see lower absorption image). In contrast, without phase jump the initial density distribution disperses (left column).}
  \end{figure}
  
  For our experiments on three-component solitons we employ a quasi-one-dimensional Bose-Einstein condensate (BEC) of $^{87}$Rb trapped in a homogeneous magnetic field.
  We realize the different components with the magnetic sublevels $\mf=0,\pm1$ of the $F=1$ hyperfine manifold.
  The soliton we are investigating is a coherent superposition of all $\mf$ fields, where $\mf=0$ features a density minimum accompanied by a phase jump.
  The bright components in $\mf=\pm1$ feature density maxima at the same position.
  In the nonlinear physics context, this type of excitation is known as dark-bright-bright soliton \cite{kevrekidis2016}.
  The fixed phase relation between the bright components allows defining the associated polarization vector.
  This is a genuine feature for solitonic excitations with at least two bright components which we control and detect in our experiment.
  
  To generate this type of nonlinear excitation, we use a spatially localized spin rotation based on the vector Stark shift \cite{marti2014} realized with a steerable laser beam (for details see \cite{sm}). This coherently transfers atoms from the initial $\mf=0$ to the bright components (see Fig.~\ref{fig:soliton_prep}(a)). The Gaussian beam profile with a root mean square (rms) radius of approximately \SI{4}{\micro\meter} leads to density distributions of the magnetic sub-states via the corresponding position-dependent Rabi coupling $\Omega(x)$ which is proportional to the modulation amplitude of the vector Stark shift and thus to the corresponding light intensity \cite{marti2014}.
  
  Simultaneously, a spatially dependent phase is imprinted which is close to the phase structure of the vector soliton, namely a phase step in the $\mf=0$ field and constant phases in $\mf=\pm1$. For this we use that a Rabi oscillation of duration $\tau$ induces field amplitudes according to $\psi_0\propto\cos(\Omega\tau)$ in $\mf=0$. Thus, rotation angles $\Omega\tau>\pi/2$ in the center of the light beam induce a region with flipped sign of $\psi_0$.
  In Fig.~\ref{fig:soliton_prep} we show the subsequent dynamics which is probed via Stern-Gerlach absorption images.
  For $\Omega\tau>\pi/2$ we find indeed that a pair of solitons is formed (Fig.~\ref{fig:soliton_prep}(b), right column) while for shorter Rabi couplings the initial density distribution disperses (Fig.~\ref{fig:soliton_prep}(b), left column).
  
  \begin{figure}
    \includegraphics[width=\linewidth]{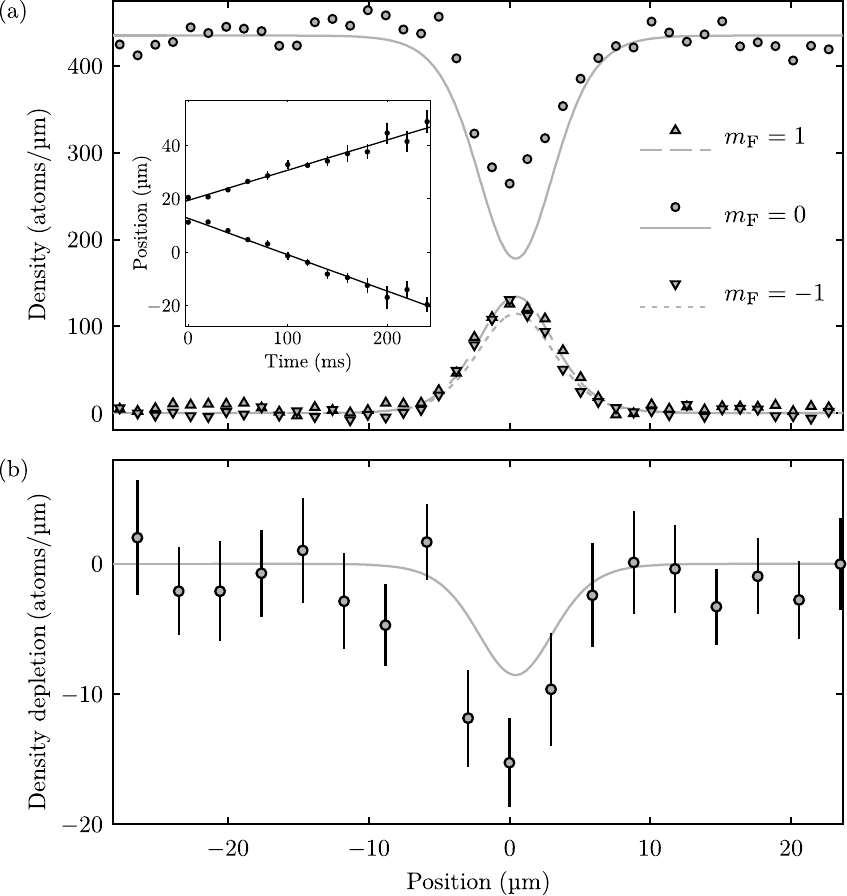}
    \caption{\label{fig:soliton_profile} Quantitative comparison with a repulsive three-component Manakov vector soliton solution.
    (a) Experimentally extracted density profiles (markers) of a single realization at $t=\SI{100}{ms}$ compared to the analytical prediction (solid lines) of Eq.~(\ref{eq:prinariWaveFct}) with independently extracted parameters from the experimental observations (see main text). The spreading of the atomic absorption signal induced by the imaging setup is taken into account by convolving the model densities with a Gaussian with rms radius of \SI{1.2}{\micro\meter} \cite{kunkel2018,sm}. The inset shows the experimentally extracted soliton positions (solid lines are linear fits).
    (b) The total density of the three-component Manakov soliton features a small depletion (solid line for parameters used in (a)) which we confirm by taking the difference between the total densities with and without solitons measured by omitting the Stern-Gerlach separation of the components and averaging over 20 realizations. All error bars mark the 1 s.d.~interval of the mean.}
  \end{figure}

  \begin{figure*}
    \includegraphics[width=\linewidth]{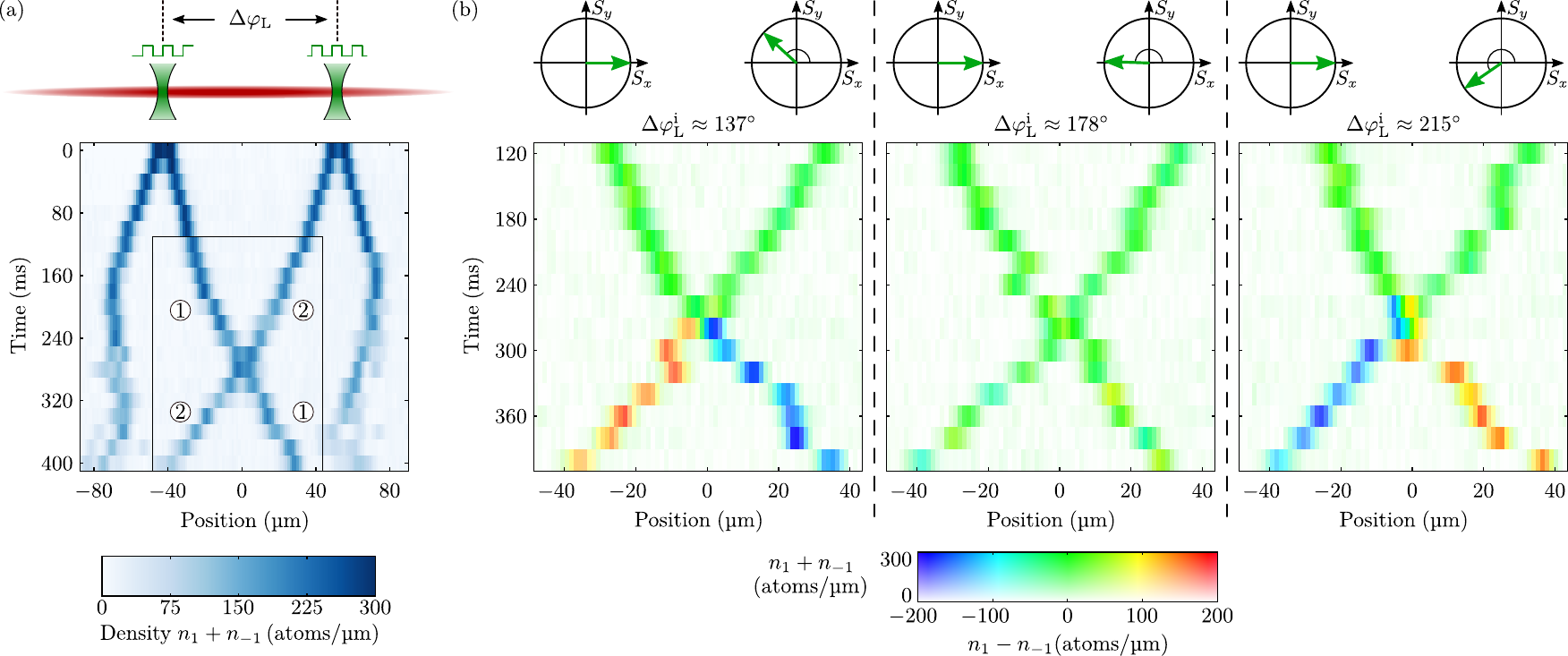}
    \caption{\label{fig:sol_coll_qual} Collision of three-component vector solitons.
    (a) We generate two pairs of solitons by applying two local spin rotations. The resulting evolution of the bright components $n_1+n_{-1}$, averaged over 6 realizations, is shown in the lower panel. Encircled numbers label the solitons.
    (b) Detailed view of the collision area marked by the rectangle in (a) for three different relative Larmor phases $\Delta\phil=\phil^{(\text{2})}-\phil^{(\text{1})}$ imprinted by corresponding phases of the laser beam modulation. The orientation of the pseudo-spin 1/2 in the $x$-$y$-plane is indicated by the green arrows. The color-coded density difference $n_1-n_{-1}$ between the two bright components of the vector soliton after collision reveals a strong dependence on the initial $\Delta\phil^\text{i}$. The saturation of the color indicates the density $n_1+n_{-1}$. We confirm that the shape of the solitons remains unaltered and that the polarization features a $\Delta\varphi_\text{L}$-dependent change after the collision.}
  \end{figure*}
  
  We confirm the preparation of three-component vector solitons by comparing our experimental observations with an appropriate analytical model.
  We expect our system to be well described by a repulsive three-component Manakov model with density interactions of equal coupling strength between all components.
  Beyond density coupling, our multi-component $^{87}$Rb Bose gas features also spin interactions, which lead to a small spin-dependent modification of the density interactions (less than one percent), and to spin changing collisions redistributing the population between the components (see \cite{sm} for details). We strongly suppress the latter process by working in a regime of positive quadratic Zeeman energy, more than 20 times larger than the spin interaction energy (i.e., deep in the polar phase \cite{kawaguchi2012}).
  The size of the solitons of $\sim\SI{6}{\micro\meter}$ is larger than the transverse extent of $\sim\SI{4}{\micro\meter}$ of the atomic cloud, therefore a one-dimensional model is adequate to describe our system.
  
  For the three-component variant of the Manakov model, it is well known that dark-bright-bright solitons exist for the system subject to non-zero boundary conditions \cite{nistazakis2008}. Recently, for solutions of the form
  \begin{gather}
    \begin{aligned} \label{eq:prinariWaveFct}
      \psi_{\pm1}(x)&=c_{\pm1}\eta\,\sin\alpha\;\text{sech}\left[\kappa\left(x-x_0\right)\right], \\
      \psi_0(x)&=e^{i\varphi_\text{S}}\left\{i\cos\alpha+\sin\alpha\,\tanh\left[\kappa\left(x-x_0\right)\right]\right\},
    \end{aligned}
  \end{gather}
  an inverse-scattering analysis has been developed in order to predict the change of their characteristics upon collision \cite{prinari2015}. In Eq.~(\ref{eq:prinariWaveFct}) the indices label the different $m_\text{F}$ states, $c_{\pm1}$ the entries of the polarization vector $\bm{c}$ associated with the bright soliton components, $\kappa$ the inverse soliton width, $x_0=\tilde{x}_0+vt$ the position of the soliton propagating with velocity $v$, and $\varphi_\text{S}$ a relative phase between $\mf=0$ and $\mf=\pm1$.
  The remaining quantities for our atomic system are given by $\eta=\sqrt{1-(\hbar^2\kappa^2/m+mv^2)/\mu}$ and $\tan\alpha=\hbar\kappa/(mv)$, where $m$ denotes the atomic mass and $\mu$ is the chemical potential of the total background density (see \cite{sm}).
  
  We parametrize the soliton polarization
  \begin{equation} \label{eq:polarization}
    \bm{c}=\frac{1}{\sqrt{2}}
    \begin{pmatrix}
      \sqrt{1+S_z}\,e^{-i\phil/2} \\
      \sqrt{1-S_z}\,e^{+i\phil/2}
    \end{pmatrix},
  \end{equation}
  motivated by the collective pseudo-spin 1/2 representation of the bright components, in terms of $S_z=(N_{+1}-N_{-1})/(N_{+1}+N_{-1})$, with $N_{\pm1}$ representing the atom numbers in the corresponding bright components. The Larmor phase $\phil$ is given by the transversal spin $S_x+iS_y=\lvert S_\perp\rvert e^{i\phil}$ with $S_x\propto\int(\psi_{+1}^*\psi_{-1}+\text{c.c.})\,dx$ and $S_y\propto\int(-i\psi_{+1}^*\psi_{-1}+\text{c.c.})\,dx$. In this language the coherence is given by the length $\lvert\bm{S}\rvert=\sqrt{S_z^2+\lvert S_\perp\rvert^2}$ and is equal to 1 in the theoretical framework of the model.
  
  For comparison with the analytical model we independently determine the polarization parameters $\phil$ and $S_z$ where we estimate $N_{\pm1}$ by summing over three times the fitted rms width. The position $x_0$ and the inverse width $\kappa$ are extracted as the mean from independent fits to the $\mf=\pm1$ components. The velocity $v$ is obtained from the position assuming linear motion of the solitons (cf.~inset of Fig.~\ref{fig:soliton_profile}(a)), and $\mu$ from the background density $n$ (see \cite{sm} for details).
  In Fig.~\ref{fig:soliton_profile}(a) we compare the individual densities $n_{0,\pm1}\propto\lvert\psi_{0,\pm1}\rvert^2$ with the solution Eq.~(\ref{eq:prinariWaveFct}) and find good quantitative agreement.
  We attribute the remaining deviations in amplitude and width of the $\mf=0$ profile to the filling up of the density minimum during time of flight for spatially separating the hyperfine levels and imaging.
  
  An additional feature of the multicomponent soliton is a maximal depletion $\delta n/n=\hbar^2\kappa^2/(m\mu)$ of the total density relative to the background density $n$.
  For our parameters we expect approximately 60 atoms to be missing in the total number of particles. This is on the order of the atomic shot noise of the total atom number over the size of the soliton. To achieve this precision we image without Stern-Gerlach separation and subtract total density profiles without solitons, each averaged over 20 realizations \footnote{This data set has been taken just before  the shut down of the lab as part of the containment procedures during the COVID-19 pandemic.}. The result in Fig.~\ref{fig:soliton_profile}(b) is close to the expectation and we find a depletion of $\sim100$ atoms which corresponds to $\delta n/n\approx0.03$ of the background density.
  
  We now turn to the study of collisions, a defining characteristic of solitons. For this we consecutively generate two soliton pairs by applying two separate local spin rotations where
  the experimental control allows modifying the soliton polarization. Here we tune the initial Larmor phase difference $\Delta\phil^\text{i}=\phil^{(\text{2})}-\phil^{(\text{1})}$ of the colliding solitons by adjusting the relative phase of the amplitude modulation of the laser beams (see Fig.~\ref{fig:sol_coll_qual}(a)).
  After approximately $t\approx\SI{260}{ms}$ the two central solitons collide without significantly changing their shape.
  However, we observe a strong variation of the outgoing soliton polarization as a function of the initial polarization difference, exemplified for three settings of $\Delta\phil^\text{i}$ shown in Fig.~\ref{fig:sol_coll_qual}(b). While for $\Delta\phil^\text{i}\approx180^\circ$ the polarization is not altered, we observe a significant change of $S_z$ for other angles.
  For the cases shown the collision redistributes the populations in $\mf=\pm1$ such that the outgoing solitons mainly contain one dominant bright component (population ratio of $\sim0.8/0.2$).
  In \cite{liu2020} both structures have been discussed as minimal energy states of the spin-1 Hamiltonian Eq.~(S1).
  
  \begin{figure}
    \includegraphics[width=\linewidth]{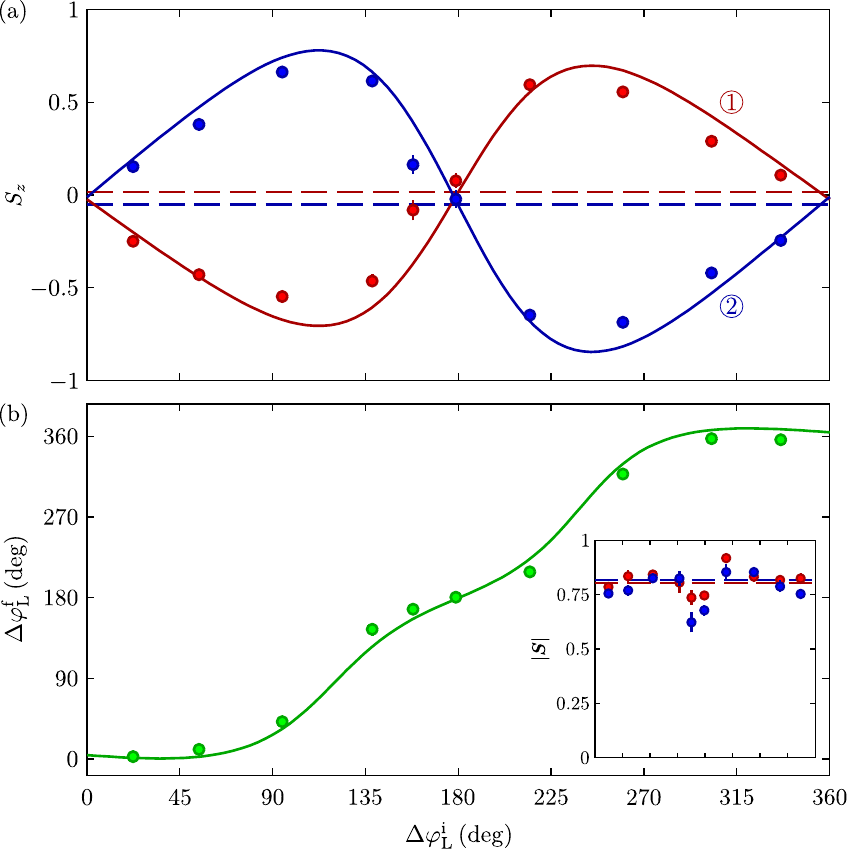}
    \caption{\label{fig:sol_coll_quant} Quantitative comparison of the experimental soliton polarization dynamics with the analytical solution. We compare the measured polarization after collision (circles, averaged over times $t=\SIrange{320}{400}{ms}$) with the predictions using independently determined model parameters (solid lines) for different $\Delta\phil^\text{i}$ measured before collision.
    (a) $S_z$ of both solitons (red and blue). As a reference the dashed lines show the experimental $S_z$ before collision averaged over times $t=\SIrange{40}{220}{ms}$ and all measured phases. We attribute the different amplitudes to the differences in initial velocities, widths and $S_z$ of the two solitons.
    (b) The measured Larmor phase difference $\Delta\phil^\text{f}$ after collision matches the analytical solution. The inset shows the experimentally measured pseudo-spin 1/2 length before (dashed line, averaged as in (a)) and after (circles) collision, revealing the conservation of coherence. The ticks on all $x$-axes correspond to the same values indicated at the bottom of (b) and all error bars indicate 1 s.d.~interval of the mean.}
  \end{figure}
  
  For further characterization we apply a detection scheme for simultaneous readout of orthogonal transversal projections of the pseudo-spin degree of freedom \cite{kunkel2019} with which we access initial and final $\Delta\phil$ as well as the transversal spin length $\lvert S_\perp\rvert$ (see \cite{sm} for the readout sequence). This allows the quantitative comparison of experimental data and theoretical predictions.
  In Fig.~\ref{fig:sol_coll_quant} we show the experimentally extracted polarization parameters $S_z$ and relative phase $\Delta\phil^\text{f}$ after collision as a function of the initial phase difference $\Delta\phil^\text{i}$.
  For the repulsive three-component Manakov model the post-collision polarizations of soliton 1 and 2 are given by
  \begin{equation}
    \begin{aligned} \label{eq:polarizationChange}
      \bm{c}_1^\text{f}=\chi\left(\bm{c}_1^\text{i}+A_{12}\left\langle\bm{c}_2^\text{i}\middle\vert\bm{c}_1^\text{i}\right\rangle\bm{c}_2^\text{i}\right), \\
      \bm{c}_2^\text{f}=\chi\left(\bm{c}_2^\text{i}+A_{21}^*\left\langle\bm{c}_1^\text{i}\middle\vert\bm{c}_2^\text{i}\right\rangle\bm{c}_1^\text{i}\right),
    \end{aligned}
  \end{equation}
  where $\langle\cdot\vert\cdot\rangle$ denotes the complex inner product, with the polarization vectors $\bm{c}^\text{i}$ and $\bm{c}^\text{f}$ before and after collision, respectively. The normalization factor $\chi$ and the coupling parameters $A_{jk}$ depend on the velocities and widths of the colliding solitons as well as on the background density (for the theoretical analysis see \cite{prinari2015}, for the connection to our experiment see \cite{sm}).
  The outgoing polarization can be seen as a superposition of the transmitted part and an admixture of the reflected part weighted with the overlap between the two polarization vectors of the incoming solitons.
  We calculate the parameters of Eq.~(\ref{eq:polarizationChange}) from the experimental quantities and find good quantitative agreement between experiment and analytical predictions (see Fig.~\ref{fig:sol_coll_quant}).
  The initial experimental asymmetries in the soliton polarization, width, and velocity cause slight amplitude differences in the post-collisional $S_z$ which are also captured by the theory.
  Our measurements reveal that the collisions conserve the pseudo-spin length $\lvert\bm{S}\rvert$ (see inset Fig.~\ref{fig:sol_coll_quant}(b)) which confirms the coherence-preserving nature of the collisions.
  
  Summarizing, we present a novel method for controlled generation of coherent multi-component solitons and verify the key features of three-component vector-solitonic propagation and interactions experimentally. We find quantitative agreement with analytical predictions of collision-induced polarization shifts in the repulsive three-component Manakov model.
  The scalability of the technique provides the means for direct generation of solitonic lattices or random soliton gases \cite{schwache1997,schmidt2012,wang2015,redor2019} with a spin degree of freedom.
  Combined with the observed long lifetime of the solitons the regime of multiple soliton collisions can be investigated.
  Notably, the collisional properties can also be described by an attractive two-component Manakov model \cite{prinari2015} for the bright components. Thus our work paves the way for the study of bright-soliton collisions in the robust environment of repulsive BECs. Combined with the decoupled spin degree of freedom this leads to long coherence times -- a new route to quantum solitons entangled in the spin degree of freedom after solitonic collision.

  \begin{acknowledgments}
     We thank Barbara Prinari for pointing out details in three-component soliton collisions and Moti Segev for elucidating the connection to the work in nonlinear optics. We acknowledge experimental assistance of Alexis Bonnin in the early stages of this project.
     This work was supported by the DFG Collaborative Research Center SFB1225 (ISOQUANT), the ERC Advanced Grant Horizon 2020 EntangleGen (Project-ID 694561), the Deutsche Forschungsgemeinschaft (DFG) under Germany's Excellence Strategy EXC-2181/1 - 390900948 (the Heidelberg STRUCTURES Excellence Cluster), and the US National Science Foundation under Grants No.~PHY-1602994 and DMS-1809074 (P.G.K.). P.G.K.~acknowledges support from the Alexander von Humboldt Foundation and the Leverhulme Trust and is grateful to the Universities of Heidelberg (Kirchhoff-Institute for Physics) and Oxford (Mathematics Institute) for their hospitality.
  \end{acknowledgments}

%  \bibliographystyle{apsrev4-2}
%  \bibliography{refs.bib}
  %

  % % % % % % % % % %  SM  % % % % % % % % % %
  \clearpage
  \onecolumngrid
  
  \setcounter{equation}{0}
  \renewcommand\theequation{S\arabic{equation}}
  \setcounter{figure}{0}
  \renewcommand{\thefigure}{S\arabic{figure}}
  
  \section{Supplemental material}
    \subsection{Spin-1 system and repulsive three-component Manakov model}
      The dynamics of the spinor Bose gas is described by the Hamiltonian \cite{kawaguchi2012,stamperkurn2013}
      \begin{gather}
        \begin{aligned} \label{eq:spinorHamiltonian}
          \hat{H} = \hat{H}_0 + \int dV \biggl[&\,: \frac{c_0}{2}\hat{n}^2 + c_1\hat{n}_0\left(\hat{n}_{+1}+\hat{n}_{-1}\right) + \frac{c_1}{2}\left(\hat{n}_{+1}-\hat{n}_{-1}\right)^2 + c_1\left(\hat{\psi}^{\dag}_{+1}\hat{\psi}^{\dag}_{-1}\hat{\psi}_{0}\hat{\psi}_{0}+h.c.\right) : \\
          &+ p\left(\hat{n}_{+1}-\hat{n}_{-1}\right) + q\left(\hat{n}_{+1}+\hat{n}_{-1}\right) \biggr]\,,
        \end{aligned}
      \end{gather}
      where $\hat{H}_0$ contains the spin-independent kinetic energy and trapping potential, $\hat{\psi}_{j}^{\dag}$ is the bosonic field creation operator of the magnetic substate $j\in \{ 0,\pm1 \}$, $\hat{n}_j=\hat{\psi}_j^\dag\hat{\psi}_j$, $\hat{n}=\sum_j\hat{n}_j$, $::$ denotes normal ordering, and $p$ and $q$ label the linear and second-order Zeeman shifts, respectively. In $^{87}$Rb the spin interaction constant $c_1$ is much smaller than the density interaction constant $c_0$ ($\lvert c_1\rvert/c_0\lesssim1/200$). Consequently, the contribution of the second and third term in the integral of Eq.~(\ref{eq:spinorHamiltonian}) is small. The fourth term, which potentially leads to redistribution between the hyperfine states, can be energetically suppressed by the second-order Zeeman shift $q$.
      
      Applying a mean-field approximation we substitute the field operators $\hat{\psi}_j$ by complex fields $\psi_j$. Due to the strong transversal confinement we approximate our system to be one-dimensional. Further, we neglect the small terms $\propto c_1$ and the phase evolutions due to $p$ and $q$ as well as the longitudinal trapping potential. Subtracting the background chemical potential $\mu$, the equation of motion reads
      \begin{equation} \label{eq:gpe}
        i\hbar\frac{\partial}{\partial t}\bm{\Psi} = -\frac{\hbar^2}{2m}\frac{\partial^2}{\partial x^2}\bm{\Psi} + c_0^{1\text{D}}\lvert\bm{\Psi}\rvert^2\bm{\Psi} - \mu\bm{\Psi}\,,
      \end{equation}
      where $\bm{\Psi}=(\psi_{+1},\psi_0,\psi_{-1})^T$ describes the wave functions of the three magnetic substates and $c_0^{1\text{D}}$ is the effective one-dimensional density interaction constant. To scale the quantities of Eq.~(\ref{eq:gpe}) to the dimensionless form of the repulsive three-component Manakov model employed in \cite{prinari2015} we use the relations
      \begin{align}
        \frac{nc_0^{1\text{D}}}{2\hbar}\,t \;&\rightarrow\;t\,, \\
        \frac{\sqrt{mnc_0^{1\text{D}}}}{\hbar}\,x \;&\rightarrow\;x\,, \\
        \frac{1}{\sqrt{n}}\bm{\Psi} \;&\rightarrow\;\bm{q}\,, \\
        \frac{\mu}{nc_0^{1\text{D}}} \;&\rightarrow\;q_0^2\,. \label{eq:q0}
      \end{align}
      With this mapping we obtain the Manakov model
      \begin{equation}
        i\frac{\partial}{\partial t}\bm{q}=-\frac{\partial^2}{\partial x^2}\bm{q}+2\left(\lvert\bm{q}\rvert^2-q_0^2\right)\bm{q}\,.
      \end{equation}

    \subsection{Solution for vector soliton collisions}
      The post-collisional polarizations of the solitons in the Manakov model are given by (see Eq.~(6.23) of \cite{prinari2015})
      \begin{equation} \label{eq:polarization_change}
        \begin{aligned}
          \bm{c}_1^\text{f} &= \chi\left(\bm{c}_1^\text{i} + A_{12}\left\langle\bm{c}_2^\text{i}\middle\vert\bm{c}_1^\text{i}\right\rangle\bm{c}_2^\text{i}\right), \\
          \bm{c}_2^\text{f} &= \chi\left(\bm{c}_2^\text{i} + A_{21}^*\left\langle\bm{c}_1^\text{i}\middle\vert\bm{c}_2^\text{i}\right\rangle\bm{c}_1^\text{i}\right), \\
        \end{aligned}
      \end{equation}
      with coupling factors
      \begin{equation}
        A_{jk} = \frac{z_j^*\left(z_k^*-z_k\right)\left(q_0^2-\lvert z_k\rvert^2\right)}{z_k\left(z_j^*-z_k^*\right)\left(q_0^2-z_j^*z_k^*\right)}
      \end{equation}
      and normalization
      \begin{equation}
        \chi = \left[1+\frac{\left(z_1^*-z_1\right)\left(z_2-z_2^*\right)\left(q_0^2-\lvert z_1\rvert^2\right)\left(q_0^2-\lvert z_2\rvert^2\right)}{\lvert z_1-z_2\rvert^2\lvert q_0^2-z_1z_2\rvert^2}\left\vert\left\langle \bm{c}_1^\text{i}\middle\vert\bm{c}_2^\text{i}\right\rangle\right\vert^2\right]^{-1/2},
      \end{equation}
      with the eigenvalues $z_j=\xi_j+i\,\nu_j$ associated with soliton $j$, where $\xi_j$ and $\nu_j$ denote dimensionless measures of velocity and inverse width, respectively, and are related to physical units by
      \begin{align}
      \xi_j &= \sqrt{\frac{m}{nc_0^{1\text{D}}}}\,v_j\,, \label{eq:xi} \\
      \nu_j &= \frac{\hbar}{\sqrt{mnc_0^{1\text{D}}}}\,\kappa_j\,. \label{eq:nu}
      \end{align}
      Note that the background density of the dark component, encoded in the dimensionless quantity $q_0$, affects the polarization change of the bright components in Eq.~(\ref{eq:polarization_change}).
      
      The parameter
      \begin{equation}
        \varphi_\text{S} = \exp\left\{i\left[\alpha-\xi_jx+(\xi_j^2-\nu_j^2)t+\theta\right]\right\}
      \end{equation}
      with phase offset $\theta$ used in Eq.~(1) absorbs all relative phase contributions between $\mf=0$ and $\mf=\pm1$ detailed in \cite{prinari2015}; for the experimental evaluation of the soliton polarization this phase evolution is irrelevant.

    \subsection{Experimental details}
      We prepare a BEC of $\sim10^5$ atoms in the state $(F,\,m_\text{F}) = (1,0)$ in an optical dipole trap of $1030\,$nm light with trapping frequencies $(\omega_\parallel,\omega_\perp) \approx 2\pi\times(1.6,170)\,$Hz in a magnetic bias field of $B=\SI{0.884}{G}$ (except for measurement of Fig.~2, where $B=\SI{1.45}{G}$), leading to $q\approx\SI{56}{Hz}\times h$ ($q\approx\SI{151}{Hz}\times h$). A magnetic field gradient along the longitudinal trap axis leads to a differential Larmor precession between two solitons. From independent measurements on the soliton evolution we determine this gradient and compensate the gradient to less than $\SI{10}{nG\per\micro\meter}$.
      
      To perform local spin rotations we utilize a laser beam at a wavelength of $\SI{790.03}{nm}$, such that the scalar Stark shift vanishes. We steer the beam by two perpendicular acousto-optical deflectors and focus it on the atoms through the imaging objective in a direction transversal to the longitudinal trap axis and the magnetic offset field. In this configuration, the circular polarization component of the laser induces a vector Stark shift which can be interpreted as a small transversal magnetic field. We use square-wave amplitude modulation at the Larmor frequency corresponding to the magnetic offset field. This fulfills the resonance condition of a Raman transfer between the magnetic sub-states. The relative Larmor phase of the solitons is adjusted by the phases of the amplitude modulations for the two spin rotations.
  
    \subsection{Experimental extraction of soliton parameters}
      Here we outline the steps for extracting the soliton parameters from the experimental data. First, we fit $\text{sech}^2$ or $\tanh^2$ functions to the different hyperfine levels independently. To account for the spatial spread of the atom signal induced by our imaging setup \cite{kunkel2018} we convolve the fit functions and analytically calculated density profiles with a Gaussian with rms radius of \SI{1.2}{\micro\meter}.
      
      We average over the fit results in $\mf=\pm1$ to extract the inverse width $\kappa$ and peak position $x_0$ in each single realization.
  
      For extracting the soliton velocity $v$ from the positions $x_0$ we apply two different methods. In Fig.~2 we calculate $v$ in each single realization from the difference between $x_0(t)$ and the average initial $x_0(t=0)$. For Figs.~3 and 4 we are only interested in the average velocity shortly before collision. Therefore, we obtain $v$ from a linear fit to the $x_0$ evolution for $t=\SIrange{160}{240}{ms}$.
  
      As described in the main text, we extract $S_z=(N_{+1}-N_{-1})/(N_{+1}+N_{-1})$ where we obtain $N_{\pm1}$ by summing the respective hyperfine density profiles over three rms widths of the soliton.
  
      The background density is the $\mf=0$ density at the soliton position when no soliton is present. For Fig.~2 we extract this value by averaging $n_0$ over the region outside the soliton where the influence of the trap curvature is still negligible. Close to the soliton collision in Figs.~3 and 4 a large region in the trap center is occupied by the two solitons. Therefore, we obtain an estimate of the background density at the soliton positions from a quadratic fit to the $n_0$ profile of the trap. For this we neglect the regions which are influenced by the solitons.
      
      From these quantities the values of $\eta$ and $\alpha$ are determined according to the equations given in the main text. For calculating $\eta$ we additionally estimate the background chemical potential. For this we apply the local density approximation and assume our trap to be cylindrically symmetric. We obtain the chemical potential
      \begin{equation}
        \mu=\omega_\perp\sqrt{\frac{mnc_0}{\pi}}
      \end{equation}
      in the radial Thomas-Fermi approximation.
      
      For an overview of the experimentally extracted soliton parameters used in Figs.~2 and 4 see Table \ref{tab:soliton_params}.
      
      The signal of the total density depletion shown in Fig.~2(b) is small. Thus, drifts of $\sim 3\,\%$ in the total atom number are expected to slightly modify the measurement results in this case. Therefore we take these into account. To reduce imaging noise we apply a fringe removal algorithm detailed in \cite{ockeloen2010} for all measurements shown in Fig.~2.
      \begin{table*}
        \centering
          \begin{tabular}{c|c|c|c}
          Quantity & Single realization in Fig.~2 & Soliton 1 in Fig.~4 & Soliton 2 in Fig.~4 \\[5pt]
          \hline & & & \\[-5pt]
          
          Velocity $v\;(\si{mm/s})$                                     & \hspace{5em}$-0.16$\hspace{5em} & \hspace{3.5em}$0.18$\hspace{3.5em} & \hspace{3.3em}$-0.21$\hspace{3.3em} \\[5pt]
          Inverse width $\kappa\;(1/\si{\micro\meter})$            & $0.31$  & $0.33$ & $0.34$ \\[5pt]
          Polarization parameter $S_z$                             & $0.08$  & $0.01$ & $-0.05$ \\[5pt]
          Background density $n\;(\text{atoms}/\si{\micro\meter})$ & $435$   & $471$  & $456$
        \end{tabular}
        \caption{\label{tab:soliton_params}
          Experimentally extracted soliton parameters for the single realization shown in Fig.~2 and the parameters used for calculating the theoretical curves in Fig.~4.
          For Fig.~4 we obtain the densities at $t=\SI{220}{ms}$ (before collision). For the analytical prediction (solid lines) we obtain $\xi_j$ and $\nu_j$ from these values and $q_0^2$ is obtained from the mean of both background densities.}
      \end{table*}
    
    \subsection{Simultaneous readout of transversal pseudo-spin projections and Larmor phase}
\begin{figure*}
        \centering
        \includegraphics[width=\textwidth]{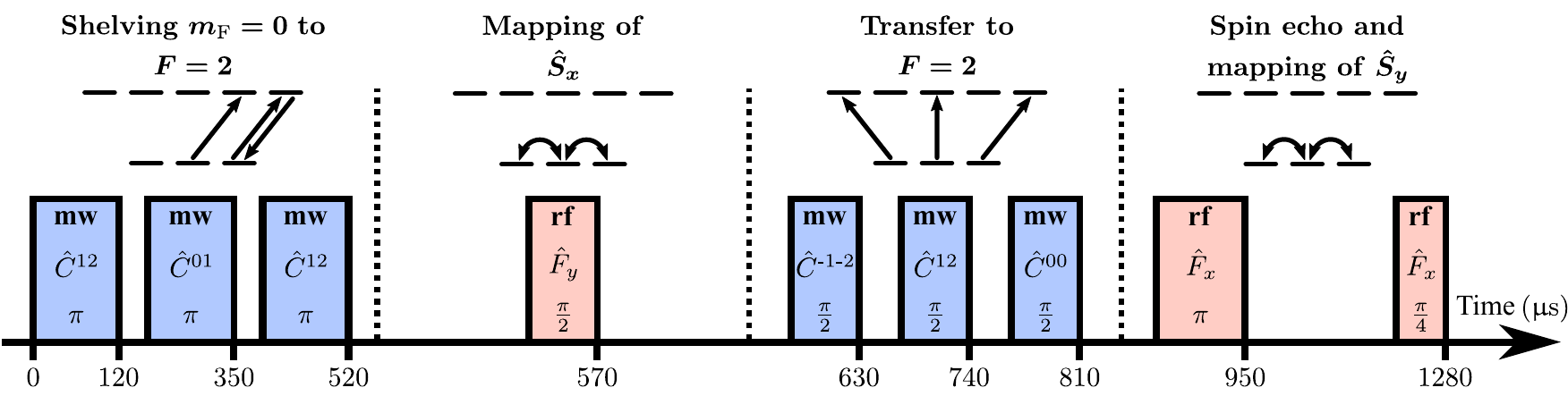}
        \caption{Pulse sequence for the simultaneous readout of $ S_x $ and $ S_y $. The level scheme depicts the coupling in the two hyperfine manifolds and the boxes contain the rotation operators together with the rotation angles.}
        \label{PulseSeq}
      \end{figure*}

      To extract the Larmor phase and the transversal pseudo-spin length of the solitons we simultaneously read out the two components $ S_x $ and $ S_y $.
      The corresponding pulse sequence is depicted in Fig.~\ref{PulseSeq}.
      To read out the transversal spin we have to access the coherence between the $ m_\text{F}=  \pm1 $ components. Therefore, we first transfer the population in the $ m_\text{F} = 0 $ component to the state $ (F,m_\text{F}) = (2,1) $, such that it does not contribute to the coherence measurement.
      Since the resonance frequency of the transfer $ (1,0)\leftrightarrow (2,1)$ is nearly the same as for the transition $ (1,1)\leftrightarrow (2,0) $, a microwave frequency (mw) pulse resonant to the first transition would also transfer atoms from the state $ (1,1) $ to the $ F=2 $ manifold. Thus, we use three mw pulses to achieve the transfer of the $ m_\text{F} = 0 $ component.
      We first apply a mw pulse to transfer all atoms from the state $ (1,1) $ to $ (2,2) $. After this the atoms in $ (1,0) $ are coupled to the $ F=2 $ manifold without disturbing the remaining populations. In a next step we employ another mw pulse to switch the populations of the states $ (2,2) $ and $ (1,1) $ back to the initial configuration.
      The mw coupling between the states $ (1,i)\leftrightarrow (2,j) $ is described by the operators
      \begin{equation}
        \hat{C}^{ij} = \frac{1}{\sqrt{2}}\left(\hat{\psi}^\dagger_{(1,i)}\hat{\psi}^{\phantom{\dagger}}_{(2,j)}+\hat{\psi}^\dagger_{(2,j)}\hat{\psi}^{\phantom{\dagger}}_{(1,i)}\right).
      \end{equation}
      
      To read out the coherence between the states $ (1,\pm1) $ we employ a radio frequency (rf) pulse corresponding to a $ \pi/2 $ spin-1 rotation which selectively couples the magnetic substates in the $ F=1 $ manifold (see \cite{kunkel2019} for details).
      Depending on the phase of the rf pulses, these rf rotations are described by the spin-1 operators
      \begin{equation}
        \begin{aligned}
          \hat{F}_x &= \frac{1}{\sqrt{2}}\,\hat{\psi}_{(1,0)}^\dagger\left(\hat{\psi}_{(1,+1)}+\hat{\psi}_{(1,-1)}\right)+\text{h.c.} \\
          \hat{F}_y &= \frac{i}{\sqrt{2}}\,\hat{\psi}_{(1,0)}^\dagger\left(\hat{\psi}_{(1,+1)}-\hat{\psi}_{(1,-1)}\right)+\text{h.c.}
        \end{aligned}
      \end{equation}
      Here, the first rf pulse sets the phase reference which we define to be a rotation described by the operator $ \hat{F}_y $. This rotation maps the $ S_x $ component of the transversal spin onto the populations of the magnetic substates in $ F=1 $.
      Before also reading out the orthogonal spin direction we use three mw pulses, coupling the states $ (1,\pm1)\leftrightarrow (2,\pm2) $ and $ (1,0) \leftrightarrow (2,0) $, to transfer on average half of the population from the $ F=1 $ manifold to $ F=2 $, where the latter serves as a storage for the expectation value of $ S_x $.
      
      Afterwards, we employ another rf pulse coupling the $ F=1 $ manifold with a relative phase shift of 90$^\circ$ such that we map the $ S_y $ component onto the population in $ F=1 $. The last rf pulse amounts to a $ \pi/4 $ spin rotation. In order to minimize the effect of magnetic field fluctuations on the readout, we use a spin echo technique corresponding to an additional rf rotation which switches the states $ (1,\pm1) $.
      After this pulse sequence, the two components of the transversal spin can be evaluated via
      \begin{equation}
        \begin{alignedat}{2}
          S_x &=  &(N_{(2,+2)}+N_{(2,-2)}-N_{(2,0)})/N_{F=2}\,, \\
          S_y &= -&(N_{(1,+1)}+N_{(1,-1)}-N_{(1,0)})/N_{F=1}\,,
        \end{alignedat}
      \end{equation}
      where $ N_{F=1,2} $ corresponds to the atom numbers measured in the respective hyperfine manifold, excluding the state $ (2,1) $. Analogously to the extraction of $S_z$ described in the main text, here the atom numbers are also summed over three times the rms width of the respective summed density profiles contributing to $N_{F=1,2}$ (this is indicated with the integrals in the definition of $S_x$ and $S_y$ in the main text).
      From this measurement we extract the Larmor phase together with the transversal spin length via
      \begin{equation}
        \left|S_\perp\right|\mathrm e^{i \varphi_\text{L}} = S_x+iS_y\,.
      \end{equation}
    
\end{document}